\documentclass[a4paper,11pt]{article}
\usepackage{pos}

\usepackage{amsmath}
\usepackage{graphicx}
 \usepackage{slashed}

\newcommand{\be}{\begin{equation}} \newcommand{\ee}{\end{equation}}
\newcommand{\ba}{\begin{array}{c}} \newcommand{\ea}{\end{array}}
\newcommand{\bea}{\begin{eqnarray}} \newcommand{\eea}{\end{eqnarray}}

\title{ChPT and lattice QCD studies of doubly charmed baryons}
\author[a,b]{Ze-Rui~Liang}
\author[a]{Jing-Yu~Yi}
\author[c,d]{Liuming~Liu}
\author*[a,e]{De-Liang~Yao}

\affiliation[a]{School of Physics and Electronics, Hunan University, 
Changsha 410082, China}
\affiliation[b]{College of Physics and Hebei Key Laboratory of Photophysics Research and Application, Hebei Normal University, 
Shijiazhuang 050024, Hebei, China}
\affiliation[c]{Institute of Modern Physics, Chinese Academy of Sciences, \\
 Lanzhou 730000, China}
\affiliation[d]{University of Chinese Academy of Sciences,
Beijing 100049, China}
\affiliation[e]{Hunan Provincial Key Laboratory of High-Energy Scale Physics and Applications, Hunan University,
410082 Changsha, China}

\emailAdd{liangzr@hebtu.edu.cn}
\emailAdd{yijingyu@hnu.edu.cn}
\emailAdd{liuming@impcas.ac.cn}
\emailAdd{yaodeliang@hnu.edu.cn}

\abstract{The scattering lengths on the interactions between the spin-$1/2$ doubly charmed baryons and Nambu-Goldstone bosons are of great importance for the investigation of the spectroscopy of heavy flavored baryons. To that end, we have conducted a systematic analysis of the low-energy dynamics of doubly charmed baryons within the frameworks of chiral perturbation theory (ChPT) and lattice quantum chromodynamics (QCD). On the one hand, the S- and P-wave scattering lengths are predicted in a manifestly relativistic baryon ChPT at leading one-loop order. On the other hand, results of the S-wave scattering lengths for four elastic scattering single channels are obtained in lattice QCD for the first time.
}

\FullConference{The 11th International Workshop on Chiral Dynamics (CD2024)\\
26-30 August 2024\\
Ruhr University Bochum, Germany\\}

\begin{document}
\maketitle

\section{Introduction}
In the quark model with four quarks of $(u, d,s,c)$, the spin-$1/2$ doubly charmed baryons (DCBs) emerge  in the $\mathbf{20}_M$-plet representation of flavour SU(4) group~\cite{Zyla:2020zbs}. Their existence is of crucial importance in classifying the heavy hadron spectroscopy. Furthermore, they provide a unique platform for exploring the non-perturbative dynamics of light quarks in the environment of two heavy quarks, which can be used to test the correctness of various theoretical models. Therefore, both theoretical and experimental investigations of the DCBs have been extensively carried out in recent years.

The search for $\Xi_{cc}^+$ was first conducted by the SELEX collaboration two decades ago~\cite{SELEX:2002wqn}. However, the reported results of $\Xi_{cc}^+$ in Refs.~\cite{SELEX:2002wqn,SELEX:2004lln} remain unconfirmed by any other experiments: FOCUS at the Tevatron proton-antiproton collider~\cite{Ratti:2003ez}, BaBar and Belle at electron-positron colliders~\cite{BaBar:2006bab,Belle:2006edu}, and LHCb at the LHC proton-proton collider~\cite{LHCb:2019gqy,LHCb:2021eaf}. 
The $\Xi_{cc}^+$ mass reported by SELEX also disagrees with theoretical predictions obtained by heavy quark effective theory~\cite{Korner:1994nh}, lattice QCD~\cite{Lewis:2001iz,Liu:2009jc,Brown:2014ena}, relativistic quark model~\cite{Ebert:2002ig}, effective potential models~\cite{Karliner:2014gca} and so on. These discrepancies have cast long-standing doubts on the existence of doubly charmed baryons. The puzzle was addressed by the discovery of $\Xi_{cc}^{++}$ at LHCb~\cite{LHCb:2017iph,LHCb:2018pcs}. Hopefully, the other two states, $\Xi_{cc}^+$ and $\Omega_{cc}^+$, will be observed in the near future as more data are being accumulated~\cite{Cerri:2018ypt,LHCb:2021rkb,LHCb:2021eaf}.

In this proceeding, investigations of the interactions between the DCBs and Goldstone bosons (GBs), both within the frameworks of baryon chiral perturbation theory (BChPT) and lattice quantum chromodynamics (QCD), are discussed. Results of scattering lengths are predicted.

\section{Scattering lengths and phase shifts from BChPT}

The Lorentz decomposition of the invariant amplitude for a given process of  the type, $\psi_{cc}(p) \phi(q) \rightarrow \psi_{cc}^\prime(p^\prime) \phi^\prime(q^\prime)$, can be written as
\begin{align}\label{eq:ABform}
\mathcal{T}_{\psi_{cc}\phi\to\psi_{cc}^\prime\phi^\prime}(s,t)=\bar{u}(p^\prime,\sigma^\prime)\left\{A(s,t)+\frac{1}{2}(\slashed{q}+\slashed{q}^\prime)B(s,t)\right\}u(p,\sigma)\ ,
\end{align}
where $\psi_{cc}\in\{\Xi_{cc}^{++},\Xi_{cc}^{+},\Omega_{cc}^{+}\}$ and $\phi\in\{\pi^\pm,\pi^0,K^\pm,K^0,\bar{K}^0,\eta\}$ denote doubly charmed baryons and Goldstone bosons, respectively. Here, $u(p,\sigma)$ ($\bar{u}(p^\prime,\sigma^\prime)$) is the spinor of the initial (final) baryon with momentum $p$ ($p^\prime$) and spin $\sigma$ ($\sigma^\prime$). There are two independent Lorentz invariant scalar products for a $2\to2$ scattering process. As usual, the commonly used Lorentz invariants are defined as
\begin{align}\label{Mandvars}
 s = \left(p + q\right)^2, \quad t = \left(p - p^\prime\right)^2\ .
\end{align}
The symbols $A$ and $B$ are scalar functions of the Mandelstam variables $s$ and $t$. Analogously to $\pi N$ scattering~\cite{Becher:2001hv,Yao:2016vbz}, one always introduces a combination of $A$ and $B$ (with $u=\sum m_i^2-s-t$), 
\begin{align}
D(s,t)=A(s,t)+\nu B(s,t)\ , \quad \nu =\frac{s-u}{2(m_{\psi_{cc}}+m_{\psi_{cc}^\prime})}\ ,
\end{align}
which is practically suitable for performing chiral expansion. For a calculation of the $\psi_{cc}$-$\phi$ scattering amplitude up to ${\cal O}(p^3)$, the following chiral effective Lagrangian is required,
\begin{eqnarray}
\mathcal{L}_{\rm eff} = \mathcal{L}_{\phi\phi}+\mathcal{L}_{\psi_{cc} \phi} \ ,\label{eq.Lag.tot}
\end{eqnarray}
with the mesonic and baryonic pieces given by~\cite{Gasser:1983yg,Gasser:1984gg,Qiu:2020omj,Liu:2023lsg}
\begin{align}
\mathcal{L}_{\phi\phi}&= \frac{F^2}{4}\langle \partial^\mu U\partial_\mu U^\dagger+(\chi U^\dagger+U\chi^\dagger)\rangle + \sum_{i=4}^8 L_i \mathcal{O}_i^{(4)}\ ,\\
\mathcal{L}_{\psi_{cc} \phi} &=\bigg[\bar{\psi}_{cc}(i\slashed{D}-m)\psi_{cc}+\frac{g}{2}\bar{\psi}_{cc}\slashed{u}\gamma_5\psi_{cc}\bigg]+ \sum_{j=1}^{7}b_j\mathcal{O}^{(2)}_j+\sum_{k=11}^{20} c_k \mathcal{O}^{(3)}_k\ .
\end{align}
In the above Lagrangians, $F$, $m$ and $g$ denote the pion decay constant, the baryon mass and the axial coupling constant in the chiral limit, in order. Besides, $L_i$ ($i=4,\cdots,8$) are $\mathcal{O}(p^4)$ mesonic low energy constants (LECs), which are determined elsewhere~\cite{Gasser:1983yg,Gasser:1984gg}. The $b_j$ ($j=1,\cdots,7$) and $c_k$ ($k=11,\cdots,20$) are $\mathcal{O}(p^2)$ and $\mathcal{O}(p^3)$ baryonic LECs, in units of GeV$^{-1}$ and GeV$^{-2}$, respectively. 

\begin{figure}[ht]
\centering
\includegraphics[width=0.6\textwidth]{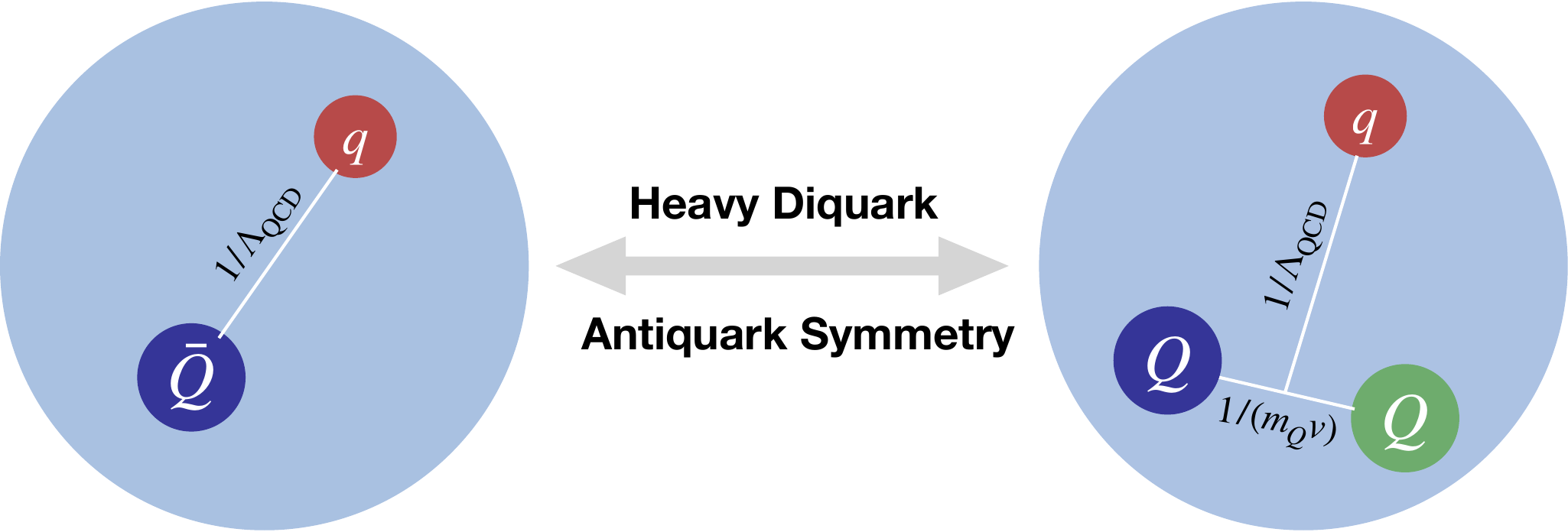}
\caption{Heavy diquark-antiquark symmetry.\label{fig.HADS} }
\end{figure}

The baryonic LECs are unknown parameters. Nevertheless, one may estimate them by imposing the so-called heavy-diquark-antiquark (HDA) symmetry~\cite{Hu:2005gf}. That is, the $cc$ di-quark system has a $\bar{3}$ representation of the SU(3) color group, which is nothing but identical to the $\bar{c}$ quark, as illustrated in Fig.~\ref{fig.HADS}. The HDA symmetry indicates that the doubly charmed baryons $\psi_{cc}$ and the charmed meson $\bar{D}$ degenerate to form a superfield. As a result, the LECs involved in $\psi_{cc}\phi$ scattering are related to the ones appearing in $\bar{D}\phi$ scattering. For a detailed discussion, the readers are referred to Appendix~D of Ref.~\cite{Liang:2023scp}. In Table~\ref{tab:lecs}, we show the values of the relevant LECs, which are obtained by using the HDA symmetry. The definition of the $\bar{D}\phi$ LECs can be found in, e.g., Refs.~\cite{Yao:2015qia,Guo:2015dha}. Their values are pinned down by fitting to lattice QCD data~\cite{Liu:2012zya}. Aspects of the LECs in the $\bar{D}\phi$ Lagrangian are comprehensively discussed in Ref.~\cite{Du:2016tgp}.
\begin{table*}[htbp]
\centering
\caption{\label{tab:lecs} Values of the LECs determined by imposing HDA symmetry.}
\renewcommand{\arraystretch}{1.2}
\renewcommand{\tabcolsep}{1.1pc}
\begin{tabular}{ c c |c c  }
\hline
\hline
\multicolumn{2}{c|}{$\psi_{cc}\phi$ scattering} & \multicolumn{2}{c}{$\bar{D}\phi$ scattering~\cite{Yao:2015qia}}\\
LEC     & value      & LEC  & value \\
\hline
$g$      & $-0.19$    & $g_0$ &{$1.095$} \\
$b_1$    & $-0.04$    & $h_0$    & $0.0172$\\
$b_2$    & $-0.11$    & $h_1$ &  $0.4266$\\
  %\hline
$b_3$    &$-1.46_{-0.46}^{+0.43}$  & $h_3$ &$5.59_{-1.96}^{+2.07}$ \\ 
$b_4$    &$0.66\pm 0.19$ & $h_2$ &$2.52_{-0.74}^{+0.73}$  \\ 
$b_5$    &$-0.17_{-0.06}^{+0.05}$ & $h_5$ &$-0.71_{-0.24}^{+0.23}$ \\
$b_6$    &$0.11\pm 0.04$ &  $h_4$ &$-0.47_{-0.17}^{+0.17}$ \\
\hline
$c_{11}$ &$-0.08_{-0.14}^{+0.21}$ & $g_2$ &$-0.16_{-0.39}^{+0.52}$  \\
$c_{12}$  &$0.08_{-0.02}^{+0.03}$ & $g_3$ &$0.08_{-0.03}^{+0.03}$  \\
$c_{20}$ &$0.49^{+0.09}_{-0.15}$ & $g_1$ &$-0.99_{-0.18}^{+0.30}$ \\
\hline
\hline
\end{tabular} 
\end{table*}

We are now in the position to make predictions of scattering lengths. For elastic scatterings, S- and P-wave scattering lengths can be written in terms of $A$ and $B$ amplitudes,
\begin{align}%\label{scattlenv2}
  a_{0+}^{(S, I)} &= \frac{m_{\psi_{cc}}}{4\pi\left(m_{\psi_{cc}} + m_{\phi}\right)} \bigg\{A^{(S, I)}(s, 0) + m_{\phi} B^{(S, I)}(s, 0)\bigg\}_{{\bf q}^2 = 0}, \notag \\
  a_{1+}^{(S, I)} &= \frac{m_{\psi_{cc}}}{6\pi\left(m_{\psi_{cc}} + m_{\phi}\right)} \bigg\{\left[\partial_{t}A^{(S, I)}(s, t)\right] + m_{\phi} \left[\partial_{t}B^{(S, I)}(s, t)\right]\bigg\}_{t = 0,\, {\bf q}^2 = 0}, \notag \\
  a_{1-}^{(S, I)} &= a_{1+}^{(S, I)} - \frac{1}{16\pi m_{\psi_{cc}}\left(m_{\psi_{cc}} + m_{\phi}\right)}\bigg\{A^{(S, I)}(s, 0) - \left(2m_{\psi_{cc}} + m_{\phi}\right) B^{(S, I)}(s, 0)\bigg\}_{{\bf q}^2 = 0}.
  \label{eq:sca.len}
\end{align}
The superscripts $S$ and $I$ stand for the quantum numbers of strangeness and isospin, respectively, while the subscripts $J^P$ represent the spin $J$ and parity $P$.
Here, $|{\bf q}|$ is the modulus of the momentum in the center-of-mass (CM) frame. The chiral expressions of the $A$ and $B$ functions are calculated up to $\mathcal{O}(p^3)$ with the Lagrangian specified in Eq.~\eqref{eq.Lag.tot} by using the extended-on-mass-shell (EOMS) scheme and heavy-baryon (HB) formalism in Refs.~\cite{Liang:2023scp} and~\cite{Meng:2018zbl}, respectively. Based on EOMS-BChPT amplitudes, numerical results of the scattering lengths are predicted and compiled in Table~\ref{tab.SL}. It is worth noting that, for future reference, the S-wave phase shifts for the elastic scattering processes, as listed in the first two columns, are plotted for the energy region near the respective lowest thresholds in Ref.~\cite{Liang:2023scp}.

\begin{table*}[htbp]
\centering
\caption{\label{tab.SL}Results of scattering lengths in BChPT. The S- and P-wave scattering lengths are in units of $\rm{fm}$ and $10^{-2}~{\rm fm}^3$, respectively.}
\renewcommand{\arraystretch}{1.2}
\renewcommand{\tabcolsep}{0.4pc}
\begin{tabular}{cc|ccc}
\hline \hline 
{$(S, I)$} &{Processes}  
&$a_{0+}$~[$J^P=\frac{1}{2}^-$]
&$a_{1+}$~[$J^P=\frac{3}{2}^+$]
&$a_{1-}$~[$J^P=\frac{1}{2}^+$]
\\
\hline
$(-2, \frac{1}{2})$ & $\Omega_{cc}\bar{K}\rightarrow \Omega_{cc}\bar{K}$     &$-0.09_{-0.13}^{+0.12}$  &$-2.47_{-2.64}^{+3.04}$    &$-0.13_{-2.64}^{+3.03}$    \\
%\hline
$(1, 1)$ & $\Xi_{cc}K\rightarrow \Xi_{cc}K$  &$-0.60\pm{0.13}$   &$-0.73_{-2.62}^{+3.02}$ &$-1.90_{-2.61}^{+3.01}$   \\
%\hline
$(1, 0)$ & $\Xi_{cc}K\rightarrow \Xi_{cc}K$  &$1.03\pm{0.19}$    &$-6.93_{-3.21}^{+2.83}$ &$-7.59_{-3.20}^{+2.82}$  \\
%\hline
$(0, \frac{3}{2})$ & $\Xi_{cc}\pi \rightarrow \Xi_{cc}\pi$ &$-0.16\pm0.02$   &$-40.6_{-2.97}^{+3.20}$    &$19.3_{-2.97}^{+3.19}$  \\
\hline
$(-1, 0)$ & $\Xi_{cc}\bar{K} \rightarrow \Xi_{cc}\bar{K}$  &$1.19_{-0.21}^{+0.22}$ &$6.19_{-5.40}^{+4.78}$  &$-3.61_{-5.37}^{+4.77}$   \\
$       $ & $\Omega_{cc} \eta \rightarrow \Omega_{cc} \eta$ &$0.42_{-0.19}^{+0.18}+0.55i$   &$0.93_{-1.96}^{+2.04}+0.01i$  &$-0.32_{-1.95}^{+2.03}+0.01i$  \\
\hline
$(-1, 1)$ & $\Omega_{cc} \pi \rightarrow \Omega_{cc} \pi$  &$-0.01\pm0.02$      &$-6.32_{-1.82}^{+1.85}$   &$-6.75_{-1.82}^{+1.85}$     \\
$       $ & $\Xi_{cc}\bar{K} \rightarrow \Xi_{cc}\bar{K}$  &$0.27_{-0.13}^{+0.13}+0.10i$   &$-4.2_{-1.21}^{+1.23}+0.01i$    &$-4.72_{-1.22}^{+1.24}+0.01i$  \\
\hline
$(0, \frac{1}{2})$ & $\Xi_{cc}\pi\rightarrow \Xi_{cc}\pi$ &$0.34\pm0.02$   &$21.9_{-3.70}^{+3.39}$     &$-104.1_{-3.70}^{+3.38}$ \\
$      $ & $\Xi_{cc}\eta\rightarrow \Xi_{cc}\eta$   &$0.06_{-0.15}^{+0.14}$
  &$-2.30_{-1.13}^{+1.13}+0.01i$    &$-3.79_{-1.14}^{+1.13}+0.01i$   \\
$      $ & $\Omega_{cc}K\rightarrow \Omega_{cc}K$   &$0.66_{-0.13}^{+0.13}+0.55i$     &$1.0_{-3.01}^{+2.67}+0.01i$    &$-2.69_{-3.00}^{+2.67}+0.01i$  \\
\hline \hline
\end{tabular}
\end{table*}

\section{Lattice QCD simulation}
As mentioned above, BChPT calculations of the scattering lengths~\cite{Meng:2018zbl, Liang:2023scp} for DCBs have utilized the LECs derived from HDA symmetry.
However, a systematic first-principle investigation using lattice QCD is still lacking. With the help of L\"uscher’s formula~\cite{Luscher:1990ux} and effective range expansion(ERE)~\cite{Bethe:1949yr, Blatt:1949zz}, S-wave interactions of four single channels, $\Xi_{cc}\pi(0,3/2)$, $\Xi_{cc}K(1,0)$, $\Xi_{cc}K(1,1)$ and $\Omega_{cc}\bar{K}(-2,1/2)$, are analyzed in this section. S-wave scattering phase shifts and scattering lengths are determined.

Four $2+1$ flavor ensembles with different volumes but the same lattice spacing at two pion masses $\sim$ $300$ MeV and $\sim 210$ MeV, provided by the CLQCD collaboration~\cite{CLQCD:2023sdb}, are used in our numerical computations (see Table~\ref{tab:ensembles}).  These configurations are generated using tadpole-improved stout smeared-clover fermion and Symanzik gauge actions. The action of the valence charm quark is the Fermilab action~\cite{El-Khadra:1996wdx}, which controls discretization errors of $\mathcal{O}(am_c)^n$. The tuning of the parameters in the Fermilab action follows the method applied in Ref.~\cite{Liu:2009jc}. To enhance the signal-to-noise ratio, we implemented the distillation smearing method~\cite{HadronSpectrum:2009krc} to calculate the quark propagators.

\begin{table}[htb]
\renewcommand{\tabcolsep}{0.7pc}
\renewcommand{\arraystretch}{1.2}
\begin{tabular}{c|c|c|c|c|c|c|c}
\hline\hline
ID&$\beta$&$a$(fm)&$a m_l$&$a m_s$&$M_\pi$(MeV)&$L^3\times T$&$N_{\rm cfgs.}$ \\
\hline
F32P30&$6.41$&$0.07746(18)$&$-0.2295$&$-0.2050$&$303.9(0.6)$&$32^3 \times 96$&$750$\\
F48P30&$6.41$&$0.07746(18)$&$-0.2295$&$-0.2050$&$304.9(0.4)$&$48^3 \times 96$&$359$ \\
F32P21&$6.41$&$0.07746(18)$&$-0.2320$&$-0.2050$&$208.1(1.9)$&$32^3 \times 64$&$459$ \\
F48P21&$6.41$&$0.07746(18)$&$-0.2320$&$-0.2050$&$207.4(0.7)$&$48^3 \times 96$&$222$ \\
\hline\hline
\end{tabular}
\caption{Summary of gauge ensemble parameters employed in this work, including ensemble identifier (ID), gauge coupling $\beta$, lattice spacing $a$, dimensionless bare quark mass parameters for light and strange quarks ($am_l$, $am_s$), pion mass $M_\pi$, lattice volume $L^3 \times T$, and number of gauge configurations $N_{\rm cfgs.}$.}
\label{tab:ensembles}
\end{table}

\subsection{Finite-volume spectrum}
The finite-volume spectrum is extracted from correlation functions of appropriately constructed operators. For single-particle states GBs and DCBs, we employ interpolating operators of the form:
\begin{align}
\mathcal{O}_{\pi^+}(x)&=\bar{d}(x)^a_{\alpha} (\gamma_5)_{\alpha \beta} u(x)^a_{\beta}, 
\quad 
\mathcal{O}_{\Xi_{cc}^{++}(ccu)}(x)=\epsilon^{ijk}P_{+}[Q_c^{i T}(x)C\gamma_5 q_u^j(x)] Q_c^k(x),  \\
\mathcal{O}_{K^+}(x)&=\bar{s}(x)^a_{\alpha } (\gamma_5)_{\alpha \beta} u(x)^a_{\beta},
\quad 
\mathcal{O}_{\Xi_{cc}^+(ccd)}(x)=\epsilon^{ijk}
P_{+}[Q_c^{i T}(x) C\gamma_5 q_d^j(x)] Q_c^k(x),
 \\
\mathcal{O}_{K^0}(x)&=\bar{s}(x)^a_{\alpha} (\gamma_5)_{\alpha \beta} d(x)^a_{\beta},\quad
\mathcal{O}_{\Omega_{cc}^{+}(ccs)}(x)=\epsilon^{ijk}
P_{+}[Q_c^{i T}(x) C\gamma_5 q_s^j(x)] Q_c^k(x),
\end{align}
where $C$ is the charge conjugation matrix and $P_+=(1+\gamma_0)/2$ is the parity projector.
As for the two-particle $\psi_{cc}\phi$ operators~\cite{Prelovsek:2016iyo} with different momenta, they can be written as
\begin{align}
\mathcal{O}_{\mathbf{p}_{\mathbf{1}}, \mathbf{p}_{\mathbf{2}}}
^{\Xi_{c c} K,\ I=0}
&=\sum_{\alpha, \mathbf{p}_{\mathbf{1}}, \mathbf{p}_{\mathbf{2}}} C_{\alpha, \mathbf{p}_{\mathbf{1}}, \mathbf{p}_{\mathbf{2}}}
\left(
\frac{1}{\sqrt{2}} \mathcal{O}_{\Xi_{c c}^{++}, \alpha}(\mathbf{p}_{\mathbf{1}}) \mathcal{O}_{K^0}(\mathbf{p}_{\mathbf{2}})-\frac{1}{\sqrt{2}} \mathcal{O}_{\Xi_{c c}^{+}, \alpha}(\mathbf{p}_{\mathbf{1}}) \mathcal{O}_{K^{+}}(\mathbf{p}_{\mathbf{2}})
\right),\\
\mathcal{O}_{\mathbf{p}_{\mathbf{1}}, \mathbf{p}_{\mathbf{2}}}
^{\Xi_{c c} K,\ I=1}
&= \sum_{\alpha, \mathbf{p}_{\mathbf{1}}, \mathbf{p}_{\mathbf{2}}} C_{\alpha, \mathbf{p}_{\mathbf{1}}, \mathbf{p}_{\mathbf{2}}}
\left(
\mathcal{O}_{\Xi_{c c}^{++}, \alpha}(\mathbf{p}_{\mathbf{1}}) \mathcal{O}_{K^+}(\mathbf{p}_{\mathbf{2}})
\right),\\
\mathcal{O}_{\mathbf{p}_{\mathbf{1}}, \mathbf{p}_{\mathbf{2}}}
^{\Omega_{c c} \bar{K},\ I=1/2}
&=\sum_{\alpha, \mathbf{p}_{\mathbf{1}}, \mathbf{p}_{\mathbf{2}}} C_{\alpha, \mathbf{p}_{\mathbf{1}}, \mathbf{p}_{\mathbf{2}}}
\left(
\mathcal{O}_{\Omega_{c c}^{+}, \alpha}(\mathbf{p}_{\mathbf{1}}) \mathcal{O}_{\bar{K}^0}(\mathbf{p}_{\mathbf{2}})
\right),
\\
\mathcal{O}_{\mathbf{p}_{\mathbf{1}}, \mathbf{p}_{\mathbf{2}}}
^{\Xi_{c c} \pi,\ I=3/2}
&= \sum_{\alpha, \mathbf{p}_{\mathbf{1}}, \mathbf{p}_{\mathbf{2}}} C_{\alpha, \mathbf{p}_{\mathbf{1}}, \mathbf{p}_{\mathbf{2}}}
\left(
\mathcal{O}_{\Xi_{c c}^{++}, \alpha}(\mathbf{p}_{\mathbf{1}}) \mathcal{O}_{\pi^+}(\mathbf{p}_{\mathbf{2}})
\right),
\end{align}
where the coefficients $C_{\alpha, \mathbf{p}_{\mathbf{1}}, \mathbf{p}_{\mathbf{2}}}$ are taken from Table I in Ref.~\cite{Xing:2022ijm}. 

The energies of the single particles and two-particle systems are then extracted from the correlation functions of the corresponding operators. 
For the two-particle systems, we employ the generalized eigenvalue problem (GEVP) method~\cite{Luscher:1990ck}. A correlation matrix is constructed as:
\begin{align}
C_{i j}(t)=\sum_{t_{\rm s r c}}\left\langle\mathcal{O}_i\left(t+t_{\rm s r c}\right) \mathcal{O}_j^{\dagger}\left(t_{\rm s r c}\right)\right\rangle.
\end{align}
The GEVP is solved with a fixed and small timeslice $t_0$ ($t_0=4$ in lattice units for this study):
\begin{align}
C(t) v^n(t)=\lambda^n(t) C\left(t_0\right) v^n(t),
\end{align}
and the eigenvalues are fitted to the two-exponential form:
\begin{align}
\lambda^n(t)=(1-\left.A_n\right) e^{-E_n\left(t-t_0\right)}+A_n e^{-E_n^{\prime}\left(t-t_0\right)},
\end{align}
where $A_n$, $E_n$, and $E_n^{\prime}$ are free parameters.

\begin{table}[ht]
\begin{center}
\renewcommand{\arraystretch}{1.2}
\begin{tabular}{l|c|r|r|r|r}
\hline\hline
&& F32P30 & F48P30 & F32P21 & F48P21 \\
\hline
{$\pi$} & $m_0~\text{[GeV]}$ & $0.3040(6)$  & $0.3049(4)$ & $0.2085(19)$  & $0.2077(7)$ \\
& $c^2$ & $1.0069(42)$ & $1.0075(22)$ & $1.0582(94)$ & $1.0157(37)$\\
\hline
$K$&$m_0~\text{[GeV]}$&$0.5230(4)$&$0.5241(3)$&$0.4917(7)$&$0.4911(3)$\\
&$c^2$&$1.0026(23)$&$1.0034(15)$&$1.0097(52)$&$1.0081(25)$\\
\hline
$\Xi_{cc}$&$m_0~\text{[GeV]}$&$3.6330(8)$&$3.6369(12)$&$3.6055(14)$&$3.6080(16)$\\
&$c^2$&$0.9915(61)$&$0.9641(186)$&$1.0056(101)$&$1.0520(218)$\\
\hline
$\Omega_{cc}$&$m_0~\text{[GeV]}$&$3.7139(6)$&$3.7179(10)$& $3.6938(8)$&$3.6995(10)$\\
&$c^2$&$0.9747(37)$&$0.9665(205)$&$0.9801(51)$&$0.9976(131)$\\
\hline\hline
\end{tabular}
\caption{Dispersion relation fit results for the single particles.\label{tab:single.particle.meff.DR}}
\end{center}
\end{table}

The masses of the single particles are determined by fitting the correlation functions to a cosh function for the mesons and an exponential form for the baryons. The results are in Table~\ref{tab:single.particle.meff.DR}. The dispersion relation $E^2 = m_0^2 + c^2p^2$ is analyzed using the lowest five momenta on lattice for each of the single particles, with fit results also presented in Table~\ref{tab:single.particle.meff.DR}. While most particle effective masses exhibit good agreement with dispersion relations, the $\Xi_{cc}$ and $\Omega_{cc}$ baryons show deviations in certain ensembles due to lattice artifacts induced by their heavy quark content.

\begin{figure}[htbp]
    \centering
   % \begin{subfigure}[b]{0.8\textwidth}
        \includegraphics[width=\textwidth]{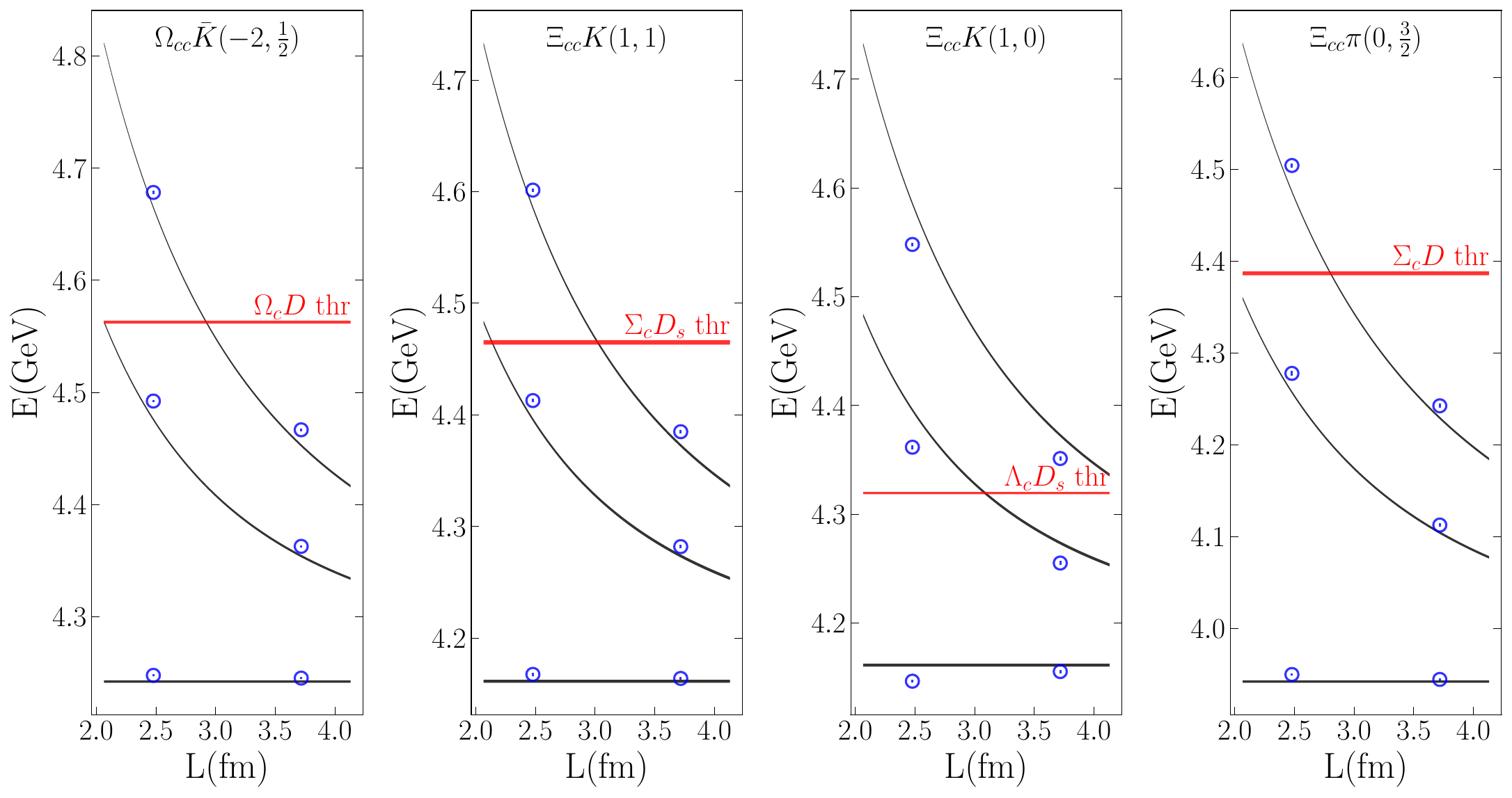}
        %\caption{}
  %  \end{subfigure}
    % \vspace{0.5cm}
   % \begin{subfigure}[b]{0.8\textwidth}
        \includegraphics[width=\textwidth]{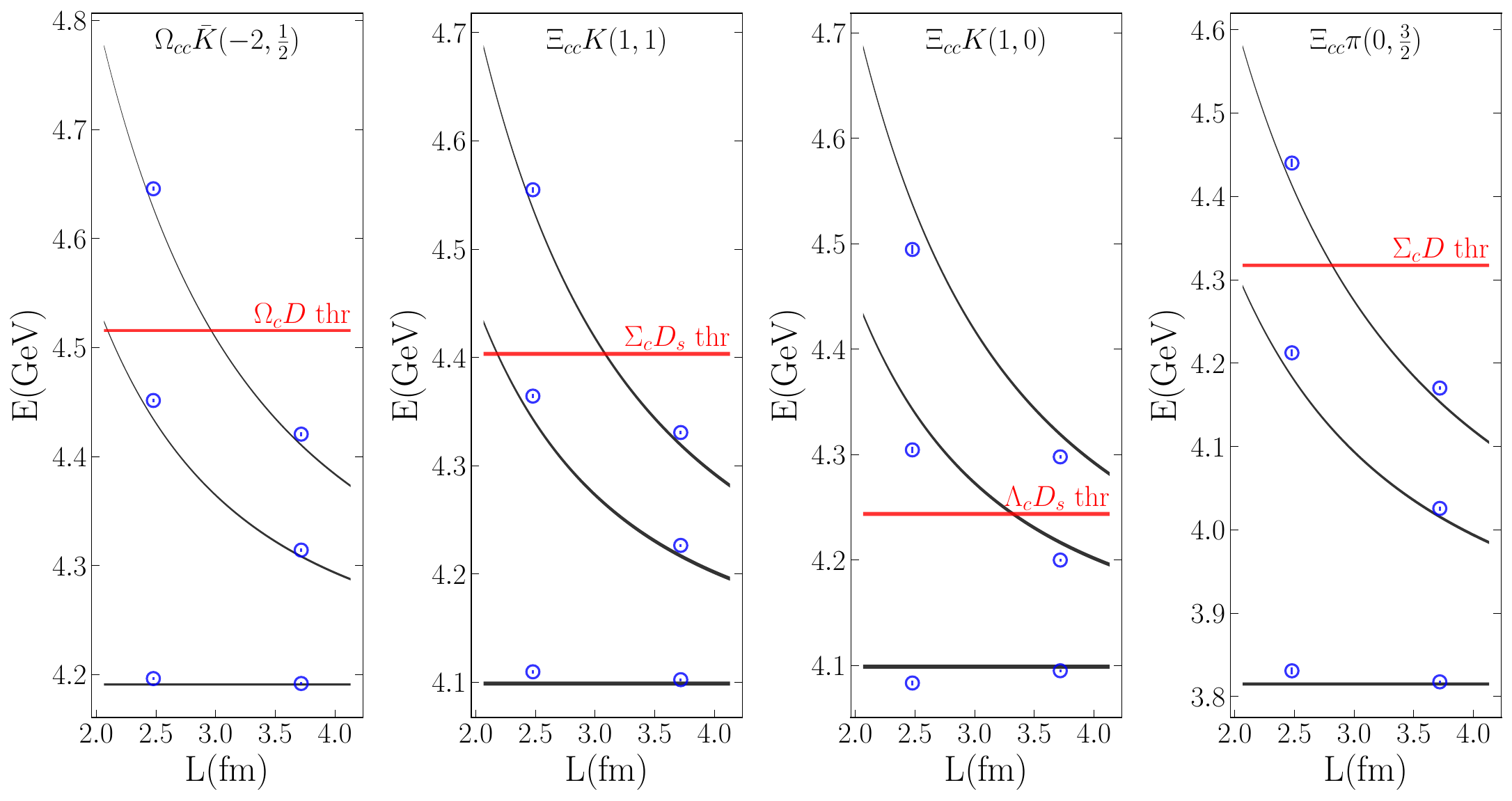}
     %   \caption{}
  %  \end{subfigure}
\caption{Energy levels of two-particle systems for the four single channels. Upper panels: $M_\pi \sim 300$ MeV; Lower panels: $M_\pi \sim 210$ MeV. The blue data points are the finite-volume energies and the black bands indicate free energies of the non-interacting threshold with different momenta.}
\label{fig:energy.levels}
\end{figure}

The finite-volume energy levels and energy shifts for scattering channels are determined from both single- and two-particle energies, as shown in Fig.~\ref{fig:energy.levels}. At spatial extent $L=48$, all channels remain close to their respective non-interacting thresholds. The $\Xi_{cc}K(1,0)$ channel displays attractive interactions, while the other three channels exhibit repulsive behavior. The observed attraction in $\Xi_{cc}K(1,0)$ suggests possible bound state or virtual state existence, requiring further investigation through scattering analysis. Notably, certain high energy levels are excluded from our analysis due to the effect of nearby channels. As illustrated in Fig.~\ref{fig:energy.levels}, the highest energy level in the $\Omega_{cc}\bar{K}$ channel exceeds the $\Omega_c D$ threshold, indicating coupling to this nearby channel. 

\subsection{Scattering analysis} 
The infinite-volume scattering parameters are extracted through the L\"uscher's finite-volume method. For a general multi-channel system, the quantization condition reads:
\bea
\operatorname{det}[\mathbf{1}+i \rho \cdot \mathbf{t} \cdot(1+i \mathbf{M})]=0,
\label{eq:Luscher.formula}
\eea
where $\rho$ denotes the phase-space factor, $t$ is the infinite-volume t-matrix, and $M$ is a matrix of known functions of scattering momentum $q=pL/2\pi$.
In the case of single-channel S-wave scattering, this reduces to the well-known relation:
\bea
p \cot \delta_0(p)=\frac{2}{L \sqrt{\pi}} \mathcal{Z}_{00}\left(1 ; q^2\right),
\label{eq:Luscher.formula.swave}
\eea
with $\delta_0$ the S-wave phase shift. The scattering parameters can be further analyzed through ERE:
\begin{equation}
p \cot \delta_{0}=\frac{1}{a_{0}}+\frac{1}{2} r_{0} p^2+\mathcal{O}\left(p^4\right),
\label{eq:ere}
\end{equation}
where $a_{0}$ and $r_{0}$ are the scattering length and effective range, respectively. 
The corresponding partial wave amplitude follows:
\begin{equation}
t \sim \frac{1}{p \cot \delta_{0}-i p}.
\label{eq:scat.amp}
\end{equation}

The scattering parameters are extracted through the implementation of L\"uscher's formula~(\ref{eq:Luscher.formula.swave}) combined with ERE~(\ref{eq:ere}). As shown in Table~\ref{tab:ScatLen}, our lattice QCD determinations of S-wave scattering lengths are in good agreement with ChPT predictions within uncertainties, even at the unphysical pion masses employed in this study.

\begin{table}[ht] 
\begin{center}
\renewcommand{\tabcolsep}{0.6pc}
\renewcommand{\arraystretch}{1.2}
\begin{tabular}{cc|cc|cc}
\hline\hline
$(S,I)$&Processes& $M_{\pi}\sim300$ MeV& $M_{\pi}\sim210$ MeV & EOMS&HB \\
\hline 
$(-2,\frac{1}{2})$&$\Omega_{cc} \bar{K} \to \Omega_{cc} \bar{K}$&$-0.161(20)$&$-0.136(12)$&$-0.09_{-0.13}^{+0.12}$ & -0.20(1) \\
\hline
$(1,1)$&$\Xi_{cc} K \to \Xi_{cc} K$ &$-0.177(23)$&$-0.212(14)$&$-0.60 \pm 0.13$ &$-0.25(1)$ \\
\hline
$(1,0)$&$\Xi_{cc} K \to \Xi_{cc} K$ &$0.63(10)$&$0.694(90)$&$1.03 \pm 0.19 $&$0.92(2)$ \\
\hline
$(0,\frac{3}{2})$&$\Xi_{cc} \pi \to \Xi_{cc} \pi$ &$-0.140(15)$&$-0.143(24)$&$-0.16 \pm 0.02 $& $-0.10(2)$ \\
\hline \hline
\end{tabular}
\caption{S-wave scattering lengths for four single channels, in units of fm. The last two columns show BChPT results using EOMS renormalization scheme~\cite{Liang:2023scp} and HB formalism~\cite{Meng:2018zbl} at physical pion mass.}
\label{tab:ScatLen}
\end{center}
\end{table}

\section{Summary and outlook}

The interactions between the DCBs and GBs have been studied both in BChPT and lattice QCD. The S- and P-wave scattering lengths for all the elastic channels are predicted with the LEC values estimated by implementing the HDA symmetry. On the other hand, the scattering lengths in the single channels, $\Xi_{cc}\pi{(0,3/2)}$, $\Xi_{cc}K{(1,0)}$, $\Xi_{cc}K{(1,1)}$, $\Omega_{cc}\bar{K}{(-2,1/2)}$, are determined at two unphysical pion masses by using four $2+1$ flavor full-QCD ensembles provided by the CLQCD collaboration. The BChPT and lattice QCD results are compared with each other, and good agreement is found, indicating the validity of HDA symmetry in the study of heavy hadrons. The obtained results provide basic inputs for future studies, which aim at performing high-precision chiral extrapolations to physical quark masses and advancing our understanding of double-heavy baryon spectroscopy.

\section{Acknowledgments}
This work is supported by Science Research Project of Hebei Education Department under Contract No. QN2025063; by National Nature Science Foundations of China (NSFC) under Contract No.~12275076, No.~11905258, No.~12335002, No.~12175279, No.~12293060, No.~12293061; by the Science Fund for Distinguished Young Scholars of Hunan Province under Grant No.~2024JJ2007; by the Fundamental Research Funds for the Central Universities under Contract No.~531118010379; by the Science Foundation of Hebei Normal University with Contract No. L2025B09.

\end{document}